# The Quantification Horizon Theory of Consciousness

T.R. Le[1]

**Abstract**

This paper engages the hard problem of consciousness (Chalmers, 1995) by examining the relationship between the neural correlates of consciousness (NCCs) and their associated qualia. The central premise is that, unlike the vast majority of systems—which have only quantifiable aspects—certain systems such as the NCCs have, in certain states, both quantifiable aspects *and* intrinsically unquantifiable aspects—the qualities of experience, or qualia. Reconciling these apparently disparate aspects within a scientific account has proven elusive, perpetuating the omission of qualia from the quantitative picture of nature established at the outset of the scientific revolution. The Quantification Horizon Theory of Consciousness (QHT) proposes a mathematical signature for this "mixed quantifiability": singularities in a complete mathematical model of the system, where quantification fails in principle. Its central conjecture is that such a model would both describe the system's physical aspects and contain singularities corresponding to its qualia. The paper also proposes where those singularities would be found: in the system's own quantitative activity. The boundary of such failures—a quantification horizon—would distinguish states of full quantifiability from states of mixed quantifiability, in which the system also has experience. Mathematics could thereby provide a signpost for qualia without reducing their intrinsic characters to quantities. The proposed correspondence paints an ontological triptych: *nothing : zero :: physical thing : value :: quale : singularity*. The paper identifies a leading candidate for the form the failure takes—a loop of comparisons, each of which can be satisfied, that cannot all be satisfied together—and considers what would establish its realization in the brain and other systems. The paper also considers implications for artificial consciousness.

---

[1]trl25@georgetown.edu, United States.

## 1 The Omission of Qualia as a Structural Consequence

The scientific revolution recast our conception of nature in mathematical terms, and it paid for that precision with an omission. Sensory qualities—colors, sounds, tastes, sensations—were left out of its picture of the world, because their felt character cannot be quantified. As Goff (2019) recounts:

> *Before Galileo philosophers took the world to be full of ... sensory qualities, things like colors, smells, tastes, and sounds. And it's hard to see how sensory qualities could be captured in the purely quantitative language of mathematics. How could an equation ever explain to someone what it's like to see red, or to taste paprika? ... Galileo solved this problem with a radical reimagining of the material world ... In [this] reimagined world, material objects have only the following characteristics: size, shape, location, [and] motion. ... The crucial point is that these characteristics can be captured in mathematics.*

The omission of sensory qualities, or *qualia*—the felt, qualitative character of experience (Nagel, 1974)—was not an oversight but a structural consequence: a model made of numbers can capture only what admits numerical characterization. If reality contains features that do not—and phenomenal experience strongly suggests it does—no refinement of quantitative models will capture their felt character.

Newton (1671/2) recognized the difficulty, writing that to explain how light produces in our minds the "phantasms of colours" is "not so easy." Leibniz (1714) made the point vivid: if we could walk inside a brain as inside a mill, we would find only pieces pushing against one another, and never anything to explain a perception. Schrödinger (1954) noted that the scientific picture cannot tell us a word about red and blue, bitter and sweet. Jackson (1982) argued that qualia have been left out of the physicalist story, Levine (1983) characterized the omission as an explanatory gap, and Chalmers (1995) named it the hard problem of consciousness.

The Quantification Horizon Theory of Consciousness (QHT) accepts the exclusion as a structural consequence: qualia are the kind of thing that a purely quantitative story cannot capture. But it asks a further question. Despite this limitation, could a mathematical model still account for the unquantifiable—not by explaining or describing it, but by marking its place, as a signpost does? QHT proposes that it can. This is the *signpost thesis*. If it is correct, and equations can mark the presence of something no quantity can convey, qualia would be a striking instance of what Wigner (1960) called the "unreasonable effectiveness" of mathematics.

What follows is a theoretical conjecture and the research programme it defines. Its philosophical core is a proposed correspondence between qualia and singularities in a complete model. The theory also proposes where such singularities would be found, and what form the failure behind them takes in our own case: a loop of comparisons that cannot all be satisfied together. Whether the brain realizes them, and how, remain open questions, though activity of a kind such loops can produce is widespread there (Section 6.2).

## 2 Foundational Commitments

Three commitments about experience underlie QHT, and they should be explicit before the theory is. If a full account of what happens in the brain could also tell us what deep blue looks like, there would be little reason to seek a special mark for qualia. These commitments give rise to the problem QHT addresses. They are starting points, not conclusions to be derived.

### *2.1 Qualia realism*

Qualia exist. There is something it is like to taste an apple, feel pain, see red. As Nagel (1974) argued, "an organism has conscious mental states if and only if there is something that it is like to be that organism." Consciousness here means phenomenal experience, whether or not the subject can speak, reflect, or recognize itself as conscious. Eliminativism, the denial that qualia exist (Churchland, 1986), is rejected here.

### *2.2 The unquantifiability of qualia*

Qualia are unquantifiable. Neural correlates can be measured, but correlates are not the experience itself. Relations between experiences can be quantified—similarity judgments, discrimination thresholds, intensity ratings—but the relata are not exhausted by those values. Qualia spaces can be mapped—opponent channels in color, for example—but structure describes relations among experiences, not what any one experience intrinsically is (Russell, 1927).

We can quantify around qualia—their causes, effects, relations, and neural correlates—but not the quale itself. This is a claim about what qualia are, not about current ignorance. Jackson's (1982, 1986) knowledge argument holds that exhaustive physical information can leave out what an experience is like. Levine's (1983) explanatory gap is the gap between structural-functional description and phenomenal character. Chalmers's hard problem generalizes the point: why should any quantitative process be accompanied by felt experience (Chalmers, 1996)? These arguments converge on the premise QHT needs: quantitative description can capture the correlates, structure, and organization of experience without capturing its intrinsic character.

A complete description of a pain, written in a book, does not hurt. By itself that proves little. A description of fire does not burn either, and no entry in any description amounts to an instance of what it describes—one application of Kant's (1781/1787) observation that existence is not a real predicate. But the two cases differ. A fire's temperature is perfectly quantifiable, and we can say how hot it is without conveying any sensation. With pain, how it feels is the very thing we seek to describe. If that feeling is part of what a pain is, every further measurement leaves the same question open.

This commitment has deep roots. Bergson (1889/1910) argued that the intensity of a sensation is itself qualitative. Russell (1927) held that physics gives structure while remaining silent about intrinsic nature. Husserl (1936/1970) diagnosed Galileo's mathematization as a substitution: a 'garb of ideas' laid over the lifeworld, in which each lived quality is exchanged for its idealized, measurable index—so that the very success of quantification is the success of a replacement, and what is replaced does not thereby vanish. Goff (2017) develops a contemporary version of this structural/intrinsic distinction.

It also has opponents. A structuralist holds that the complete pattern of relations among colors exhausts what there is to color experience. A physicalist may explain the apparent gap by the different concepts we use to think about a brain state and to recognize the experience from within. QHT's reply turns on the distinction between *determining* an experience and *characterizing* it: the physical state of a brain might determine exactly which quality a person experiences even though no description of that state could convey the quality. Readers who hold that phenomenal character is not intrinsically beyond quantitative characterization will reject QHT—but as a substantive philosophical disagreement, not an undefended stipulation.

### *2.3 Reality has both quantifiable and phenomenal aspects*

Reality has two kinds of aspects: physical aspects that are quantifiable (mass, charge, spin, position, momentum) and phenomenal aspects that are unquantifiable (the redness of red, the painfulness of pain). They are aspects of a single reality, the way a surface has both curvature and texture. This places QHT within the dual-aspect monist tradition (Atmanspacher, 2014).

A gyroscope's particles, fields and motion all admit quantitative description, and if, as we ordinarily suppose, the gyroscope has no experience, those quantities could in principle exhaust what there is to describe. It is a system of *full quantifiability*. The neural activity of a person seeing deep blue admits quantitative description too. But the premise of this paper is that the experience of deep blue is a further aspect of the very system whose activity we measure. Having both measurable properties and a quality whose intrinsic character no quantity can convey, that system is one of *mixed quantifiability*.

Our mathematical models are quantitative maps. They capture quantifiable aspects with extraordinary precision but are structurally incapable of capturing the unquantifiable—not because the models are insufficiently refined, but because assigning values is what quantitative models do, and phenomenal aspects are what values cannot capture. The question becomes: what happens in the map at the points where it meets features of the territory that it cannot represent numerically?

### *2.4 Two kinds of question*

The hard problem conflates questions of fundamentally different kinds (Majeed, 2016). Category A concerns the structure of consciousness: where qualia fit in the physicalist framework; why they are ineffable, private, subjective, unified, and causally efficacious; which systems have them. Category B concerns the intrinsic nature of experience. It comprises the palette question (which physical structures correspond to which qualitative characters?), the qualitative character question (why does experience have these specific characters rather than others?), and the existence question (why does reality have unquantifiable aspects at all?).

QHT proposes a law for when a system has experience, and asks what that law could explain about the structure and unity of experience. Particular pairings between physical states and felt qualities would require further laws, and even then we could ask why a state has the character it does, and why anything is felt in the first place. Answering the questions of Category A would nonetheless constitute enormous progress, even with the deeper aspects of the hard problem unresolved.

## 3 The Core Proposal

QHT rests on an ontological proposal about the structure of reality and its relationship to quantitative description: the *ontological triptych*. Reality, the theory proposes, presents three cases—non-being and two modes of being—each with a distinctive signature in our formal models of the world:

*nothing : zero :: physical thing : value :: quale : singularity*

Where there is nothing, a model assigns zero. Where there is something quantifiable—whether captured by a real number, a complex number, a tensor, or any other well-defined mathematical object—a model assigns a value. Where there is something unquantifiable, the model degenerates: it produces a singularity. The first two cases are familiar; four centuries of physics operate in the domain of values and zero. QHT's claim is that the third has been hiding in plain sight—that singularities are the mathematical signature of a fundamental ontological category: qualia. The singularity marks a quantification horizon—the boundary beyond which quantitative description cannot reach—and qualia are what lie behind it.

### *3.1 The bridge principle*

The triptych turns on a difference in what a model can supply: at some states a value, at others none. QHT conjectures that, in a complete model, this formal difference corresponds exactly to a difference in whether the system has experience. This is the *bridge principle*.

> *A physical system has a phenomenal aspect in a given state if and only if a complete model of the system is singular in that state: its quantification fails there in principle.*

A *complete model* does three things. It describes the physical facts about the system. It indicates the states in which the system has an experience, even though no equation could convey what the experience is like. And it derives that indication from the physical facts, which presupposes that those facts determine when experience occurs; a label added whenever a listener reported a tone would simply hand the model its answer.

*Singular* is used in an extended sense, which can be made exact. A *requirement* specifies, for each state $x$ of a system, conditions that some quantity must satisfy there; let $R(x)$ be the set of values that satisfy them. The state is *regular* for the requirement if $R(x)$ has a member and *singular* if it is empty; a model is singular in any state that is singular for one of its requirements. To fail in principle is to fail in this way: the value is not hard to obtain; there is none to obtain. A formula that divides by zero is the familiar case, but nothing need diverge, and every other quantity in the model can retain a definite, finite value. The failure itself can be described exactly, even measured; what has no value is the quantity required. Singular states can occupy an entire region, and a quantification horizon is its boundary.

The quale is on the territory side; the singularity is its signpost on the map side. A singularity is therefore not a hole in reality, any more than a map projection that fails at a pole puts a hole in the globe.

The commitments of Section 2 do not deductively prove the bridge, but they constrain the alternatives. If qualia are real (Section 2.1), they are not absences, so zero is the wrong signature. If they are unquantifiable (Section 2.2), no value captures them. And if they are aspects of the same physical states the model describes (Section 2.3), a complete model cannot simply leave them out, as a map of curvature leaves out texture. What remains is a place inside the domain where the apparatus returns no value: a singularity, the most natural structural candidate for marking something real that cannot be represented as a value.

Two claims must be kept apart, because they carry different weight. That a complete model marks the states in which a quality is present, so far as mathematics allows, is part of what 'complete' means. That the mark takes the form of a singularity is the conjecture: nothing in mathematics dictates that the marked states form the singular set. Nor has anyone constructed a complete model of the neural correlates of consciousness (NCCs); it may prove impossible to do so.

### *3.2 Why singularities? The structural parallel*

Qualia and singularities share a logical shape, which is what suggests that one is the mathematical signpost of the other. Everything around a quale can be quantified (its causes, its effects, its relations), and only at the quale itself, at what red is like, does quantification fail. A singularity reproduces that shape within the formalism: it can be located and characterized exactly, yet what the equations register there is a limit to quantification itself.

An obvious objection sharpens the point. A sign need not resemble what it signifies: the word 'pain' denotes pain without hurting. Numbers can be signs in the same way. Nothing prevents us from calling the look of deep blue 7 and light blue 31. But a number can stand for something in only two ways. It can name it, by convention, and then any other number would have served: nothing about 7, neither its size nor its place between 6 and 8, answers to anything in the look of blue. Or it can measure it, and then its arithmetic carries the meaning: heavier than, twice as long, midway between (Krantz et al., 1971). Arithmetic expresses relations, so what a measure can report is how one quality stands to others, the structure that Section 2.2 grants can be quantified. For what a quality is in itself only the first way remains, and a number used that way is a word written in digits: it indexes the quality and conveys nothing of it.

A singularity is not a sign of that kind. No one assigns it: where a complete model fails is fixed by the system it describes, and no convention could move it. But that is true of every feature such a model derives. Why, then, could some ordinary feature of a model not mark a quale just as well? Take a person looking at a plain patch of deep blue. On a natural rival to QHT, the experience consists in, or accompanies, the successful activity by which the visual system presents the color, and a scientist who identified that activity might correctly predict every occasion on which blue is seen.

For prediction, that would suffice. But QHT also seeks a reason for the choice of correlate. Everything required to distinguish blue from green can be described in terms of the responses the two colors elicit and the uses made of them, and nothing about success in that task points to a quality that no quantity can express. A singularity does. Qualities were left out of the scientific picture because, on the premise adopted here, their intrinsic character cannot be quantified; a place where the equations themselves establish a limit to quantification is therefore the natural place to look for their sign. The choice of correlate is guided by the very reason the qualities were excluded, and that rationale is available even before any such limit has been found in a brain. A number can correspond to a quality only by fiat; a singularity would correspond to it by shape.

The limit that excluded the qualities and the limit a singularity marks are not the same. A quality is something present that no value conveys; a singular state is one in which a value that was called for does not exist. What they share is the shape with which this section began. The bridge conjectures that the second is how the first appears in a complete model; the likeness recommends that conjecture and does not prove it. Section 4.2 gives the shape an exact form.

### *3.3 Where to look: the location hypothesis*

Where in the mathematics should the sign be sought? Not every omission or breakdown will do. A selective model omits features by design: a topographic map does not register color, and no singularity results. An observer can just as easily force an equation to become undefined at an arbitrary point. Even a genuine physical conflict falls short: three magnets can have couplings that no arrangement satisfies, although none of the magnets measures anything. What would make a failed quantification the system's own?

Experience always belongs to someone, so the failure that marks it should be the subject's own. A brain, unlike a magnet, does its own quantifying: when a person reaches for a ball, the visual system estimates its position relative to the body, and that estimate serves the person to whom the ball appears. That is reason to look for the limit in the system's own quantitative activity. This is the *location hypothesis*.

> *In a complete model of a physical system, the singularities that mark phenomenal states occur precisely where the system's own quantification fails in principle.*

Two conditions give ‘the system’s own’ its content. First, the system must itself be determining the quantity: some of its signals are put to use in fixing it. What counts as a signal and its use is a matter for an account of physical computation, and QHT assumes one on which what a system computes is an objective matter of its physics, not of an observer’s interpretation (Anderson and Piccinini, 2024). The magnets fail this condition. Second, the requirement must be set by those signals and that use alone, with no appeal to what the signals are about or to anyone’s purpose. A device that cannot do what its designer intended has not thereby failed at anything of its own.

A drawing of a Penrose triangle shows what such a failure could look like. Each corner appears as an ordinary joint between two solid bars, yet no solid object could have all three joints as drawn (Penrose, 1992). If the viewer’s visual system requires a single solid object with those joints, the relations the viewer actually sees make the requirement impossible to fulfill. The drawing shows why such a calculation would fail; whether the viewer’s brain undertakes it is a separate matter. And a plain patch of blue presents no such apparent conflict. Whether the brain nevertheless meets a quantitative limit in seeing it is the harder question, to which Section 6.2 returns.

The hypothesis rests on a premise that should be stated: a complete model of a system that quantifies must carry a term for each quantity the system is determining. Where the system’s requirement can be met, the term has a value; where it cannot, the model has none to assign and is singular. Without this premise a complete model could be regular throughout, as a physicist’s model of the same system is, merely noting that a result was lacking. The premise is a choice about what completeness demands, not a result. With it, half of the hypothesis follows: a complete model is singular wherever the system’s own quantification fails. That it is singular nowhere else remains conjecture.

Neither hypothesis is delivered by mathematics alone: a formalism registers only its own breakdown, and that the breakdown marks something is the identification itself. Taken together, they state the theory in one sentence: a system has experience exactly when its own quantification fails in principle.

### *3.4 What would count as evidence*

Two lines of inquiry are independent. A person who sees deep blue knows the experience from within; a neuroscientist can study the same brain from outside and ask what its activity makes possible. The conjecture would gain support if the two converged: if a model developed to explain some independently studied function of the brain proved singular in just those states in which the person has an experience. No one would have inserted a mark to fit the person’s reports; the equations would have arrived there unprompted. Whether they do is a question for investigation (Sections 6 and 8).

### *3.5 The epistemic status of the bridge*

QHT proposes a psychophysical law in the sense envisaged by Chalmers (1995, 1996). The arguments above motivate singularities as the structural candidate for such a law; they do not establish that the correspondence is real. If QHT is wrong, a quality could accompany a perfectly regular feature of brain activity, or experience might leave no recognizable mathematical trace at all.

A deletion test makes the situation vivid. Strike 'water' from chemistry and every equation stands. Strike 'quale' from the examples of the next section and the mathematics likewise survives: it still describes a limit to what a particular quantitative operation can accomplish. Phenomenality enters only through the identification. But water was identified with $H_2O$ by matching its independently known properties to those of the substance chemistry describes, and both sides of that identification were open to measurement. Here only one side is. Scientists can measure the activity in the NCCs; the person who sees blue knows the quality from within.

Section 4 therefore has a limited aim: to state precisely the form that a failure of the required kind could take. The correspondence with experience is conjectured on the strength of the considerations above and remains to be established empirically.

## 4 How Quantification Fails: The Loop That Does Not Close

The bridge speaks of quantification that fails in principle, and the location hypothesis says whose. Neither says what form such a failure takes. This section proposes one, a loop of comparisons that does not close; says when such a loop is a system's own; describes what it looks like from outside; and takes up the other objections the proposal invites.

### *4.1 When comparisons run in a circle*

Every measurement begins as a comparison: of a length with a rod, of one light with its neighbor, of a signal now with the signal a moment ago. To quantify is to turn comparisons into magnitudes, one number for each thing compared, and the step usually goes so smoothly that it passes unnoticed. It can fail even when every comparison is accurate.

Ask which is strongest: rock, paper or scissors. Every contest between two of them has a known winner, and no fact about the three is hidden. Still the question has no answer, because the wins run in a circle. Three dice can be numbered so that each beats the next five times in nine (cf. Gardner, 1970), and then even the odds are known exactly. Now consider a device whose task is to give each die a rating, a single number, so that whichever die wins more often has the higher rating. Its records are accurate and complete. What cannot be had, even in principle, is the ratings: any three numbers have an order, and the wins have none. Equal ratings are no way out: they call the dice evenly matched, and each beats the next.

The device has not come up short in the ordinary way. A flat map that shows two floors of a station as a single point cannot say which floor a traveler has reached; but there something is missing, and an altimeter would supply it. Here nothing is missing. No further fact about the dice would help, because there are no ratings to be found. In the terms of Section 3.1 the device's state is singular for its requirement: the set of ratings that meet it is empty.

### *4.2 One form*

The dice show a form that recurs wherever quantification fails with nothing missing: a circuit of comparisons, each of which can be satisfied, that cannot all be satisfied together. The parts are consistent; the whole they are taken to describe cannot exist. Go once around, and you return to a different answer from the one you left with: in that sense the loop does not close.

A loop that does not close need not go on forever. That it does not close means that no answer exists, not that the asking never stops. The asking stops when the question changes: the dice are put away, the gaze moves on, a standard is reset. What such a loop cannot do is end by succeeding. Its failure lasts as long as the same comparisons stay in force, and in a brain that may be brief (Section 7.3).

Physicists call such a circuit *frustrated*, after magnets coupled so that no arrangement satisfies every pair (Toulouse, 1977), like the three magnets of Section 3.3. The Penrose triangle of that section is the same circuit in depth: carried around its three corners, the depths of the bars never return to where they began (Penrose, 1992; Ghrist and Cooperband, 2025). The liar's sentence, which says of itself that it is false, is the circuit shrunk to a point (Abramsky et al., 2015). Hofstadter (2007) called loops of this shape strange, and looked for the self in them (Section 9.2).

The form is not one illustration among several. Where pieces of information agree wherever they overlap, and their overlaps form no circuit, a whole that fits them all always exists (Vorob'ev, 1962). And finitely many comparisons can be honored by numbers exactly when no chain of them returns to its starting point, which is where the theory of measurement begins (Krantz et al., 1971). In these settings, if nothing is missing and no two demands conflict outright, a loop is the only way to fail.

Such a state is singular in the sense of Section 3.1, and the word fits in its older sense too. A walker who circles a tower finds its compass bearing a full turn from where it began, and the circuit encloses the one place, the tower itself, where no bearing exists. A loop that does not close is how a singularity looks from around it; and around a quale, Section 2.2 observed, is where quantification succeeds.

This gives the shape of Section 3.2 an exact form. Qualities were set aside because their intrinsic character admits no magnitude, however fully their relations are measured. In a frustrated loop every relation is measured, and the magnitudes the relations were meant to yield do not exist. The two predicaments differ as that section said. What they share is the shape, and in neither would a further measurement help.

QHT conjectures that the likeness is no accident. This is the *loop hypothesis*.

> *The failures of quantification that mark a system's phenomenal states are, in our own case, frustrated loops: circuits of the system's own comparisons, each of which can be satisfied, that cannot all be satisfied together.*

It is the most specific of the theory's three claims and the most exposed. The other two do not depend on it: together they say that a system has experience exactly when its own quantification fails in principle (Section 3.3), whatever form the failure takes. A frustrated loop is one such form, and it marks experience in any system whose own it is; the loop hypothesis adds that this is the form behind the experience of persons. Should a person's experience prove to be marked by a failure of another form, such as two demands colliding on one signal, the loop hypothesis would fall and the other two would stand (Section 8.2).

### *4.3 Whose loop?*

Nothing described so far is conscious, and QHT does not say otherwise. The game is not, and the dice are not. They have the form, but the loop lies in what might be compared, and nothing is doing the comparing. The rating device does compare, and still the failure is not its own: the circuit lies in the dice, and the demand for ratings is its maker's, which the second condition of Section 3.3 excludes.

That condition allows no appeal to purpose or to what a signal is about, so a requirement must be read from a system's own signals and the use it makes of them. What a system revises does not settle the matter. A device that adjusts its ratings after every contest may be seeking only the ratings that fit its records best; for the three dice such ratings exist, and a device content with them lacks nothing. The requirement shows in what is done with the result. Suppose a device goes on revising its ratings until every contest on record bears them out, and releases them for use only then. Read as a requirement of the device's own, that condition calls for ratings that satisfy every comparison at once, whatever the signals are called and whatever the device is for. With these dice there are none to release: on that reading the device is on a loop, and the failure is its own. Such loops can arise wherever a system measures against a standard that its own readings have set.

This is how a loop realizes the third case of the triptych (Section 3). The system is determining a quantity: the magnitudes that its comparisons are to bear out. A complete model of the system must carry a term for that quantity (Section 3.3). Zero would be the right entry if nothing were being determined, and a value if the comparisons could all be honored. When they run in a circle, neither will do. Something is being determined, so zero is wrong; and every comparison is already on record, so no further fact could supply a value. There is none. The model is therefore singular at that place, in the sense of Section 3.1, and the failure is one of principle: nothing in the system has gone wrong, and nothing is missing. By the bridge, a singularity of this kind marks a phenomenal state.

Not every device that cycles is on such a loop. One that only switches on the outcome of a comparison assigns no magnitude: each comparison it makes has an outcome, and it lacks nothing, however long it cycles. An on–off thermostat is of this kind (Section 6.1), and so is a ring of switches that never settles: like the magnets of Section 3.3, it has the form of a loop and measures nothing.

Nor is the test whether a circuit settles. Where every unit drives a signal of its own toward a level that depends smoothly on the other signals, and signals and levels alike are held between fixed limits, there is always a setting at which every unit has what it is driving toward, by Brouwer's fixed-point theorem. That is as true of a circuit on a loop as of any other: the device with the dice may come to such a balance, its revisions cancelling and the same ratings put to the test again and again, with contests still against them and nothing released. What marks the loop is that at every setting, the balance included, some comparison to which the device's own operation holds its ratings goes unmet. A circuit in which nothing is held to such agreement is not on a loop, whether it settles or cycles. Whether a given device is assigning magnitudes or only switching, and what use it makes of them, is part of the problem flagged below.

If a physical system is on a loop that cannot close, QHT predicts experience in that state, however simple the system. Section 6 faces that consequence directly. Two things remain open. Whether loops of this kind occur in the brain is an empirical matter. And each example stipulated the system's signals and operations. Reading them from the physics of an arbitrary system is a problem that QHT inherits from the theory of physical computation and has not solved, and the verdicts of Section 6 depend on it.

Behind both lies a question that stipulation does not remove. In the device with the dice every operation is carried out: each contest is checked, the checks are combined, and the gate stays shut. To say that the device requires ratings it cannot have, and not merely that a check has returned its answer, is to read the condition on release as a requirement of the device's own. The location hypothesis depends on that reading. QHT offers a candidate for what licenses it, stated over a system's operations alone and without appeal to purpose. A condition is a requirement when some operation of the system accepts only a value that meets it, and the system goes on producing candidates for that operation until one does; it is a check when the outcome is passed on and the operation proceeds either way. The device with the dice does the first; a switch or a thermostat does the second. Nothing in this asks the rest of the system to wait: its other operations go on, and may draw on the unfinished work, as a person goes on acting and reporting while a loop fails to close (Section 6.1). Whether the candidate is right is an open problem distinct from the reading of operations from physics, and every verdict on a device in this paper, for experience or against it, is conditional on it.

### *4.4 The view from outside*

What a loop that cannot close looks like from outside depends on what the system does while its requirement goes unmet. Where an unmet comparison must move some signal, the signals cannot settle; where it need not, they can stand still. Taking turns is no way out, since comparisons honored one at a time have not been honored together.

Whether such activity also breaks detailed balance is a further matter. A system obeys *detailed balance* when, around every closed sequence of its states, the tendencies to change cancel (Kolmogorov, 1936); where they do not, they run in a circle, like the wins of Section 4.1. A circuit on a loop can swing between two states and back, or mend its comparisons from either side at random, and run no more one way than the other. That the activity carrying a loop departs from detailed balance is therefore a further hypothesis, which QHT makes about the circuits proposed for brains (Section 6.2), and not a consequence of the loop hypothesis. Nor would the departure be the mark of a loop: a river and a candle flame are held far from equilibrium, and neither compares anything.

### *4.5 Would a finer description dissolve the failure?*

It might seem that the failure exists only for want of detail. But where nothing is missing, detail has nothing to add: a scientist who knew every face of every die could only explain more exactly why no ratings exist. The detail must be complete, however. If a system in fact uses further signals to settle what its comparisons leave open, those signals belong in the model too, and an apparent failure may come from leaving part of the system's own activity out.

Nor is the missing value itself the quality. A scientist could measure every neural signal available to a person seeing red and establish exactly where the visual system's own comparisons fail to close. On QHT that finding would mark the presence of a quality. The failure is the sign, not the thing signified. Would still finer measurements then disclose what red looks like? A more detailed map is still a map. On the premise of Section 2.2, the look of red is known through experience.

### *4.6 Where is the unquantifiable?*

If atoms and neurons are each fully quantifiable, how can the system they compose possess an unquantifiable aspect, and where is it? On QHT the look of blue belongs to the person whose visual system is organized and active in a particular way. The model describes the activity of the parts and the relations that make them a system; its singularity marks the state in which the person sees blue. To ask which neuron possesses the look of blue is to seek a feature of the organized whole in one of its parts. Nor need anything be added to the parts: on QHT, a change in organization alone could carry a system from full to mixed quantifiability, and a scientist could trace that change down to the last neuron. A strip of paper shows how. Joined end to end, it has two sides; joined with a half-twist, it has one. Nothing was added to the paper, and the twist is at no particular place, since a cut anywhere removes it. Like the dice, the strip compares nothing and experiences nothing; it shows only that a whole can differ where no part does.

### *4.7 The quantification horizon*

Where a system's states vary continuously, the boundary of the region its singular states occupy is a *quantification horizon*. For a system that changes in discrete steps, we can instead identify the transitions from a regular state to a singular one and back. In either case the physical system must tell us which states are neighbors or which transitions are possible.

The rating device gives a small example. Suppose it asks for less than fixed ratings: magnitudes that may vary from moment to moment, provided they do not depend on which contest is being held. The dice can supply these, since on every roll each die shows a number. But on any one occasion three magnitudes have an order, so of the outcomes '*A* above *B*', '*B* above *C*' and '*C* above *A*' at most two can hold. However the magnitudes vary, the proportions of occasions on which each outcome holds must therefore satisfy

$$P(A > B) + P(B > C) + P(C > A) \leq 2.$$

Rock, paper and scissors give three. Whenever the exact probabilities sum to more than two, the things compared cannot have such magnitudes, even momentary ones. When the three probabilities are equal and exceed one half, such magnitudes can be found for every common value up to two thirds, and for none above: in that family of states, two thirds is the horizon. Boole (1862) called restrictions of this kind "conditions of possible experience," using experience in the sense of observation; they recur in Specker's parable of the overprotective seer (Liang, Spekkens and Wiseman, 2011) and in the study of Bell inequalities and contextuality (Abramsky and Brandenburger, 2011).

Not every singularity signifies, however, and general relativity shows how a physically significant limit is told from a mathematical artifact. Three of its lessons carry over.

First, a singularity can arise from an idealization, or mark the limit of a theory's applicability (Batterman, 2011; Butterfield, 2011). The familiar thermodynamic singularity at a phase transition appears only in the limit of an infinitely large system; a singularity proposed as a sign of experience in a brain would have to arise in the finite system itself. But singularities need not be artifacts of this kind. Penrose (1965) showed that, under certain conditions of gravitational collapse, geodesic incompleteness—a light ray's path ending, after a finite stretch, with no continuation within the spacetime—can follow from the equations without assuming a perfectly symmetrical collapse. On the usual view such a singularity marks the place where general relativity itself gives out, and a complete model would then have none there. The conjecture of Section 3.3, that a complete model is singular nowhere but at a system's own failures, includes that expectation, and a fuller theory that kept such a singularity would count against it.

Second, a singularity can be an artifact of coordinates. The usual Schwarzschild coordinates break down at the event horizon of a black hole, while other coordinates cover it smoothly and preserve the physical boundary beyond which light cannot escape to infinity (Carroll, 1997, §7). For QHT the corresponding result is elementary. Under a change of description that is invertible on states and on values, each set $R(x)$ is carried to its counterpart, so no such change can create a singular state or remove one. The horizon is preserved as well if the change and its inverse preserve which states are near one another or, for a system that changes in steps, which transitions are possible. Merging a state that admits an answer with one that does not, by contrast, can conceal a failure.

Third, a limit can be objective even though it depends on how a problem is posed, as a *Cauchy horizon* bounds what the data on a chosen initial surface determine (Carroll, 1997, §4). In QHT the system's own use of its signals poses the problem; once that use is specified, an outside observer can establish whether the required result exists. What it takes to specify that use is the open problem of Section 4.3.

The event horizon also gives the theory its name. Light emitted inside one cannot reach a distant observer outside, yet what lies inside is physically real; the analogy concerns only the boundary of what can be reached. In QHT a singularity can be located from outside the system, but the felt character of the state lies beyond the reach of quantitative description. The person seeing blue has access to something that a complete set of measurements could only mark.

## 5 The Hallmarks of Qualia

Any adequate theory of consciousness must accommodate five hallmarks of qualia. The taste of an apple has an ineffable character that neither numbers nor words can fully convey; it is private; it belongs to a particular subject; it forms part of one experience; and it can influence what the subject does next. For the first three, QHT follows a single line of thought: the mark is open to any observer, while what it marks belongs to the system itself. The loop hypothesis of Section 4 brings unity under the same account. Causal efficacy asks more of the theory.

### *5.1 Ineffability*

No description fully conveys the taste of an apple. A scientist could follow every physical change that takes place when someone bites into one and still, never having tasted an apple, not know the taste. The remedy would be a slice, not a further measurement.

That the taste cannot be conveyed in quantities is one of QHT's starting points (Section 2.2). What the theory adds is a way for this limit to leave a trace. The scientist could identify what the taster's sensory system quantifies and show exactly where and why its quantification gives out. That account could be mathematically exact, and by the bridge it would establish that an experience is occurring. It would be precise about the failure and silent about the taste. Ineffability, on this account, is the horizon seen from outside.

### *5.2 Privacy*

No one can literally feel another person's headache. Why are qualia private when their physical substrate is not?

A scientist could measure the firing rates of the sufferer's neurons and locate the proposed horizon, all from what Nagel (1986) called the view from nowhere. But however precise the measurements become, they describe a state of someone's brain, and the sufferer stands in a different relation to that state: the relation of undergoing it. The singularity lies on the map and the headache in the territory (Section 3.1). A map can be copied and passed from hand to hand, and so the mark is open to any observer. The territory cannot: there is one of it, and it is the sufferer. To hold the map of a headache, however exact, is not to be the thing it maps. A scientist whose own head ached in just the same way would know that kind of pain firsthand, and would still be feeling a different headache. QHT thus gives the experience a place that is open to inspection while preserving the distinction between inspecting a state and being in it.

### *5.3 Subjectivity*

Every experience is from some point of view. We feel a headache in our own heads and see a cup on the table in front of us. Privacy concerns who can feel the ache; subjectivity concerns why it is anyone's in the first place.

The location hypothesis gives a physical way to ask whose point of view is at issue. A viewer's brain estimates where the cup lies relative to the viewer's body. When the body moves, the estimate changes, and the changed estimate guides the next reach. If quantification fails within that activity, the limit already belongs to a particular system: the one whose movements supply those signals and whose actions use them. We can describe that whole process from outside without any estimate of where the cup lies relative to us.

Use, not content, settles ownership (Sections 3.3 and 4.3). Two circuits could carry the same readings at a given instant while using them differently, and the bridge attributes experience to the system whose use it is: the one that registers changes in the body and guides its actions.

### *5.4 Unity*

We experience "a red ball rolling," with color, shape and motion all belonging to the ball. The bridge can tell us that something is felt, but not why it is the moving ball that looks red.

The loop hypothesis offers a start. The failure of a loop belongs to the circuit as a whole and to no part of it: cut any one link, and what remains can be satisfied. Features whose comparisons lie on one loop would share that failure, and Section 7 proposes that such sharing determines which features appear together. The account makes a prediction: changing the physical grouping while preserving the individual features should change which object looks red.

Binding the features of one object still leaves a larger question: what makes the rolling ball and the table across which it rolls one experience for one subject? The same proposal would answer with a larger loop that takes in both (Section 7.1). Shared failures, and not what the person can report, would then mark what belongs to one experience: some of what a person experiences may be unavailable for report on a particular occasion (Block, 2011).

### *5.5 Causal efficacy*

A hand pulls back from a hot surface, and the natural explanation is that it hurt. Yet a physiologist might trace the physical causes of the movement in full. What causal role remains for the pain?

QHT locates the pain in the system that acts, without adding another force to move the hand. A critic may grant that the neural activity caused the withdrawal and still insist that the pain merely accompanied it. Would the hand have moved just as it did had nothing been felt? The question has two readings.

On the first, every physical event is held fixed, together with the bridge. The same brain then passes through the same singular states, so removing all experience while keeping that very course of events is impossible. On the second, only the movement of the hand is held fixed. A different physical route, such as a reflex, could then produce the same withdrawal, which would tell us little about what caused this one.

This addresses the easier part of Kim's (1998) exclusion problem. The pain is no rival to the physical cause; it belongs to the very physical state that helps bring the movement about. Even where its quantification fails, the brain occupies a definite state and its activity continues, and the bridge fixes whether that state has a phenomenal aspect. Whether the way the state feels does explanatory work is a further question. Hoel (2017) shows how grouping microscopic states together can sometimes give a more informative account of cause and effect. Grouping brain states by how they feel might do the same here, but only given a law identifying which states are painful and an argument for why the way they feel helps explain the movement.

## 6 Which Systems Are Conscious?

### *6.1 The criterion*

How widely should we expect to find experience? The bridge answers in both directions. In a state where quantification succeeds throughout a complete model, the system has no experience. Where quantification fails in principle in the state the system actually occupies, the bridge predicts experience, however simple the system. The location hypothesis says whose quantification is at issue: the limit must arise in some quantitative activity of the system itself.

Complexity is therefore neither required nor sufficient. What matters is whether a system depends on a quantity it cannot have. A thermostat switches the heating on below a set temperature and off above it, and its readings suffice for the one distinction it uses. A physicist could ask for much more, down to the speed of each air molecule in the room, but nothing in the thermostat's operation calls for it. A horizon would require a distinction that the device's own activity uses and cannot make.

Applying the criterion to a person seeing deep blue means establishing which parts of the brain take part in the relevant activity, on what timescale, and whether the comparisons those parts carry out can all be satisfied together.

One brain may hold more than one such system. A failure belongs to the circuit on the loop (Section 4.3), and to nothing larger merely because it lies in the same skull, so the bridge applies to each circuit, or structure of joined circuits (Section 7.1), separately. The person whose experience we judge by report and action is the system on whose loops those reports and actions depend (Section 5.3). Should that system's loops all close in dreamless sleep while loops elsewhere in the brain do not, QHT predicts no experience for the sleeper, and, for each circuit whose loop does not close, experience of the barest kind (Section 6.4) that is not the sleeper's.

### *6.2 Why look in brains?*

In brains, both sides of the conjecture can be examined: we know in our own case that some brain states are accompanied by experience, and we can study from outside what neural signals represent and how the brain uses them.

Some models of vision describe how the brain estimates what gave rise to the light reaching the eyes (Lee and Mumford, 2003), and estimates about pairs of features may contribute to an estimate of the whole figure (Parr et al., 2019), as in the rating device of Section 4.1. Neurons in visual cortex can respond differently to the same edge depending on which side of it a figure occupies (Zhou et al., 2000): to see the outline of a white cup against a dark wall as part of the cup, the visual system must keep track of which surface each edge belongs to. What happens when different edges suggest incompatible outlines?

One circuit might choose the most likely figure even when some edges suggest another. A second might use the edges to locate the outline of a single figure that fits them all. The two would agree whenever the edges agree. When the edges conflict, a best estimate may still exist for the first, even if the second has no such outline to locate. The recordings establish sensitivity to the side of a figure. They leave open which use the brain makes of those signals, and on QHT that is the question that matters.

Two further facts point to brains. Their raw material is comparison: sensory neurons report contrasts, each scaled by the activity around it (Carandini and Heeger, 2012), and magnitudes such as the lightness of a surface must be assembled from them. And human judgment sometimes shows the form openly. Shepard (1964) built a ring of tones, each heard as higher than the one before, that produced "a complete breakdown of transitivity in judgments of relative pitch," and human choices have shown contextuality of the kind studied in physics (Cervantes and Dzhafarov, 2018), which Aerts, Broekaert and Gabora (2000) proposed, more broadly, as the crux of consciousness. Whether experience depends on such circles is another matter.

Conflicting edges are occasional, and so are circular scales; experience is not. A plain blue wall is seen continuously, with nothing in view to reconcile. If the loop hypothesis is right, loops of the brain's own fail to close for as long as anything is seen, and the theory owes a candidate for what they are. The scene is the wrong place to look. Self-reference has the right profile. The comparisons at issue are among the brain's own signals, made against standards the brain itself supplies: the activity around each neuron, the neuron's own recent readings and, on predictive accounts, the brain's forecasts of its own sensory states, which it acts to bring about (Friston, 2010). These are loops in the sense of Section 4.3, and as they are usually modeled they close: a standard catches up with what it measures, and a forecast settles on its best fit, which is no failure. Nothing yet shows why any of them should be held to an agreement that cannot be had when there is nothing in view to reconcile, and that is the main open challenge for the loop hypothesis. If some are, a plain wall engages a great many of them, and wherever the comparisons on one cannot all be satisfied, the failure lasts as long as the wall is seen. The failures, on this candidate, would be many and local. That any loop fails would account for there being experience at all; which loops are engaged, and how they are joined, for what is seen (Section 7). Whether any brain is so organized is an open question.

The wiring this would take has long been known to be possible in nervous nets. McCulloch (1945) showed that a small net can be so wired that its preferences run in a circle, with no common scale of values to fit them, and in models of macaque cortex, frustrated loops keep activity from settling into one pattern (Gollo and Breakspear, 2014). Restless activity is not itself the sign. A pair of cells in which one excites the other and is inhibited in return, or a group of neurons that take turns without a winner (Rabinovich and Varona, 2018), may never settle, and yet nothing in such a circuit, as modeled, is held to an agreement that cannot be had (Section 4.3). A loop of the kind QHT proposes is one whose comparisons the brain's own operation holds to agreement and which cannot all be honored (Section 7.2).

QHT's further hypothesis is that the activity carrying such loops breaks detailed balance (Section 4.4). The anatomy allows it: neural interactions are seldom reciprocal, and physicists count networks of neurons among the systems governed by no single quantity (Fruchart et al., 2021). The break can be measured, and at the scale of the whole brain it has been: activity there departs from balance slightly at rest and strongly under physical and cognitive effort, and a model traces the departure to small asymmetries among interacting elements (Lynn et al., 2021). Two studies report that it lies nearer equilibrium in sleep and under anesthesia than in wakefulness (Sanz Perl et al., 2021; de la Fuente et al., 2023), and under deep anesthesia it is captured by a few recurring states (Wilsenach et al., 2025). These findings show less than they might seem to. The measures are coarse, and a heart would pass them (Nartallo-Kaluarachchi et al., 2026); they follow effort as well as wakefulness; and both studies count ketamine among the states of reduced consciousness, although experience can persist under ketamine when responses cease (Section 8.1). They establish only that dynamics of the kind this hypothesis expects pervade the working brain. Nor should such activity be expected to vanish when a person's experience does, since a brain holds more circuits than those that make up the person (Section 6.1). The search for loops can begin where the brain turns comparisons into magnitudes.

### *6.3 Machines and artificial consciousness*

A language model can write movingly about happiness. Even knowing how it produces those words would leave open whether it has ever felt happy (Shanahan, 2024; Chalmers, 2023). A physical sign of experience, once established in the brain, would give us something to look for in a machine.

It must be sought in the machine itself, not in what the machine computes. A computer might simulate a brain and correctly identify every state in which the brain's quantification fails, while completing every calculation it undertakes. To decide whether it also realizes the process it simulates, we must trace what its signals do in the machine (Anderson and Piccinini, 2024).

Tracing them shows where the line falls. Every program runs on hardware, so in a thin sense any loop a program describes is physical. That is not the sense required. The test of Section 4.3, with the proviso stated there, is put to the machine as it operates: does its own use of its signals call for magnitudes that every one of its comparisons bears out, and do those comparisons run in a circle? A machine of which both are true meets the test whatever it is made of, and whether or not a clock times its steps. A machine that only works out what such a system would do does not meet it: it can prove that the loop will not close, or predict its course by other means, while holding nothing of its own to an agreement that cannot be had. Between the two lies a general-purpose computer that steps through each comparison in turn. Whether such a computer, so programmed, has that requirement itself, in the sense of Section 4.3, is the problem flagged there, and QHT does not yet settle it.

A device that philosophers call counterpredictive makes the difference concrete (Scriven, 1965; Rummens and Cuypers, 2010). From a description of the wiring of a machine with two lamps, a computer can prove that the machine cannot correctly predict which lamp it will light. Nothing fails in that computation. Now suppose the computer forms a forecast of which lamp it will itself light, and feeds that forecast to a switch wired to light the other. A lamp lights on every trial, and the forecasting circuit can use each error to revise its next prediction. Yet whichever lamp it predicts, its own forecast must be wrong. Every component works exactly as designed. The computer is now in the predicament that it could only describe before. That alone is not a failure of its own quantification: a bit flipped at each mismatch is a switch, and a ring of switches lacks nothing (Section 4.3). The failure would be the computer's own if its operation held magnitudes to an agreement that cannot be had, as the rating device of Section 4.3 holds its ratings.

If a machine's own quantification failed in this way, QHT would predict experience in that state, whatever the machine was made of. Behavior alone cannot show it. Two machines might describe the same scene in the same words while differing in whether their own vision could settle a distinction they used. An engineer could establish that physical difference by tracing their circuits, though using it as evidence of a difference in what the machines felt would require independent grounds for the bridge (Doerig et al., 2019).

The failure belongs to the circuit on the loop (Section 6.1): the oscillator that paces a computer is a separate system from anything the computer runs, and what a machine says bears on its experience only if its words depend on the circuit that fails. Present-day foundation models, language models among them, produce their outputs in passes that run from input to output and end. In such a pass each value is computed once from those before it; feeding the output back as the next input adds to a sequence and adjusts nothing in it. Every value the pass calls for is supplied, then, and none is withheld until some agreement is reached. Where that description holds, no requirement of the model's own goes unmet, and on the same proviso QHT predicts that it has no experience, whatever it says of itself. The verdict concerns how such models work, not what they can describe: a system whose own operation called for magnitudes that its comparisons, running in a circle, could not supply would be another matter, whatever it was made of.

The first machine deliberately designed to have an experience might therefore be quite modest. A further ambition would be to learn what a machine's world might feel like to the machine. Knowing the physical counterparts of familiar qualities, and the laws governing their changes, might teach us how to bring about an unfamiliar one. Even then, the designer would have to undergo that state to know how it feels.

### *6.4 Selective, not panpsychist*

QHT is explicitly non-panpsychist, yet it allows experience in a simple machine while denying it to most of the physical world. The two verdicts have one source. A rock has a mass, a temperature and a composition, and a geologist may measure all three without the rock ever quantifying any of them. Chalmers (1995, 1996) distinguished the proposal that information generally has a phenomenal aspect from the proposal that only certain kinds do; QHT gives the latter a definite form. In Wallace's (2011) office building the "whisper of sourceless ventilation" is always there. Experience, on QHT, can cease altogether when a system returns to a regular state; in a brain, that holds of each system separately (Section 6.1).

Here QHT parts company with panpsychist versions of Russellian monism (Russell, 1927; Goff, 2017; Mørch, 2019). It expects most physical systems to have only quantifiable aspects, and it allows the same matter to belong to a conscious system in one state and an unconscious system in another, according to whether the system's own quantification reaches a limit there.

The price of giving experience a physical criterion is accepting the unfamiliar cases that meet it, and the price is not peculiar to QHT: Herzog, Esfeld and Gerstner (2007) argue that, for each of the leading models they consider, some very small network meets the stated conditions. An analog controller may track a changing signal, and a neuromorphic chip may feed its outputs back into its inputs, indefinitely and with every quantity they use remaining available; feedback alone is not frustration. Neither is cycling: a thermostat, a ring of switches or a natural oscillator can cycle without end and lack nothing (Section 4.3). But a device on a loop that cannot close meets the criterion, as a machine that only works out what one would do does not: the rating device of Section 4.3, which goes on revising its ratings and passes them on only when every contest bears them out, and any chip whose operation calls for magnitudes that its comparisons, running in a circle, cannot supply. In each case, on the proviso of Section 4.3, QHT predicts experience, and on the proposal of Section 7 it is of the barest kind, without memory, reflection, pleasure, pain or a unified world. How often nature realizes even this much, and what it takes to realize a world like ours, are separate questions (Section 7).

# 7 From Singularities to Rich Experience

The bridge is a yes-or-no criterion, and experience is nothing like a yes or a no. Seeing a red ball roll across a table is a single experience with distinguishable parts: its color can deepen and its motion slow while the table remains in view. How could a singularity account for a field so rich in detail and so continuous in time? The answer suggested here begins with the fact that a singularity need not be an isolated point. Several failures can form a connected structure, and the structure can persist as its parts change. What follows goes beyond the bridge, and is offered as a proposal throughout.

## *7.1 The binding problem*

The binding problem is sharpest when more than one object is in view. Take a red square beside a blue circle. Their separation, their order from left to right and the angles of the square all admit quantitative description; the colors themselves stand in relations of similarity and opposition, which mathematical approaches to experience study (Kleiner, 2020; Tsuchiya and Saigo, 2021). Yet knowing which colors and shapes are present still leaves open which figure looks red.

Exchange the colors of the two figures. All four features remain, and the scene looks different. Following Bayne's (2010) emphasis on total experience, QHT treats the square's color as an aspect of the whole scene. A physical counterpart to binding must therefore have a structure that differs between the two arrangements even though the same four features are present in both.

Frustrated loops can supply such a structure, because loops share links. Where two loops have a comparison in common, neither can be closed without regard to the other, and a scientist could establish what they share by tracing the signals that both use. And since the failure of a loop belongs to the circuit as a whole (Section 5.4), comparisons that lie on one loop share one failure.

For a single moment of seeing, such a model yields a graph. Each point stands for a frustrated loop, and a line joins two loops that share a comparison. The proposal is that the groups formed by these lines correspond to groups of features in experience. Exchanging the colors of the square and the circle, while keeping both colors and both shapes, would then require a change in the grouping, and a larger loop would have to bring both figures into the same subject's field. The loop hypothesis does not specify which loops belong together; this is one proposal for how they do.

### *7.2 The richness of experience*

How can a single condition support the intricacy of human experience? Because the condition can hold across a region in which much else varies. The red of the square can deepen while its edges stay still; the circle can move while keeping its blue. In the proposed picture, some parts of a complex singular structure vary independently while others continue to hold the features of each figure together. The form itself has room to spare. Among $N$ things there are $N(N-1)/2$ pairs to compare. Where each comparison reports by how much one thing exceeds another, only $N-1$ of them are free if all are to derive from magnitudes; the remaining $(N-1)(N-2)/2$ are independent ways not to close (Jiang et al., 2011). For a hundred things that is 4,851 against 99. Most of what such a system of comparisons can register is not a magnitude. Organization at several scales would add further room for variation (Section 7.5).

### *7.3 Continuity*

Experience feels smooth and continuous, yet individual neurons fire in brief events. A singular structure could persist while the particular neurons participating in it changed: one group remains active as a second joins it, then gives way while the second continues. If the organization that marks a heard sound persists through these successive changes, the listener hears one continuing note. The proposal makes the duration and overlap of the organized activity matter. Change comes as readily as persistence. The loops are posed anew with each movement of the eyes and each adjustment of the standards against which neurons compare, so the field can be renewed as fast as comparisons are made.

### *7.4 Structure, anchors and the palette of experience*

If qualia are unquantifiable, how can experience exhibit measurable structure? Because what is quantifiable is the relationship between qualia—their relative positions, intervals, and oppositions—while what is unquantifiable is the intrinsic character at each point. In QHT's terms, a singularity can mark the presence of a quality even though the relations among singular states are open to quantitative description.

Those relations suggest an approach to the palette question of Section 2.4. A mark that says only that a value is missing might seem too poor to tell one quality from another. But a failure is not one thing. Values differ in size; failures differ in form, in which comparisons a loop joins and which it shares with others (Sections 7.1 and 7.2). Different processes would fail in different ways, and different failures could mark different qualities. Small changes within one family of singular states might correspond to fine differences among shades, and a change in the form of a singularity to a more abrupt switch in what an observer sees. Suppose two shades of blue look more alike than either looks like yellow, and a model independently identifies two singular states more similar in a relevant physical respect than either is to a third. The proposal that such physical and experienced relations correspond is called *structural fidelity*. Related proposals include Chalmers's (1995) structural coherence, Kleiner's (2020) experience spaces and Tsuchiya and Saigo's (2021) relational account.

Resemblances alone may leave several ways to pair qualities with physical states. What narrows them is an *anchor*: an established pairing of one quality with one state. A person looking at deep blue can pick out the quality as "this shade," while measurements identify its physical counterpart.

Imagine, as a toy case, that four qualities and four physical states each form the corners of a square. Eight matchings preserve which corners are next to each other: four rotations and four reflections. Anchor one quality to one state and two remain; anchor an adjacent quality as well and only one is left. (Nothing here assumes that color experience has this geometry.) In general, if a one-to-one matching preserves the specified relations in both directions, an alternative can differ from it only by a *symmetry*, a rearrangement that preserves those same relations (cf. Palmer, 1999). Once the anchors rule out every rearrangement that would change a pairing, the matching is unique.

Laws of change could extend a matching beyond the anchored cases. Given an anchor and an independently justified rule for each change, we could follow a sequence of changes to a further state. Consistency then becomes a test: two routes to the same complete physical state, including its relevant history and context, must agree on the experience there. Composition imposes a similar check: changing the square's color must agree with changing that aspect of the whole scene.

Not every physical relation will serve. Those used for a matching must reflect how the failure arises, survive changes in units or notation, and be identified independently of the experiences they are meant to predict; the proposed pairings could then be tested on further cases. A physical feature shared by states that feel different is too coarse to identify the quality, even if it correctly marks its presence. And extending a matching from one person to another requires further grounds (Kleiner, 2020).

Such a law could determine which quality occurs without conveying what it feels like. But once some pairings and the rules connecting them were established, a separate unexplained choice would no longer be needed for every shade or change of scene.

### *7.5 Degrees of consciousness*

Whether a state is singular is a yes-or-no question, yet a scene can contain a faint outline or a wealth of sharply articulated detail. On the account sketched here, the difference lies in how much singular structure is active, how much of it belongs to one coherent field, and how many levels of organization that field sustains. Each can vary while experience continues: a patterned surface could lose its visible detail while remaining present as a single patch.

Degrees of consciousness can then be studied without assuming that every kind of degree lies on the same scale. At a specified physical scale one could examine the density of singular events through time, the proportion of the active structure belonging to one connected whole, or the depth at which local features are grouped into larger ones. Which changes yield a more detailed field, which a more unified one, and which alter the intensity of a particular quality? Greater density alone need not make an experience richer. Across all these variations, the bridge marks where experience is present.

## 8 Testability

### *8.1 Levels of prediction*

What finding could speak for or against QHT? That depends on which layer of the proposal is at issue. The bridge predicts a singularity in a complete model wherever a system has an experience. The location hypothesis directs us to the system's own quantitative activity, and the loop hypothesis to comparisons there that cannot close. A neural model must then identify what the brain quantifies and where its quantification fails in principle. Short of such a model, one thing can be measured now. If the loops are carried by circuits of the kind proposed in Section 6.2, the activity that carries them should depart from detailed balance; that is a further hypothesis about those circuits, not a consequence of the loop hypothesis (Section 4.4). A law of qualitative change would go furthest, predicting which differences among brain states correspond to differences in experience.

Ordinarily a light appears, a person sees it, and the person responds, and a neural change accompanying all three might track any one of them. Sarasso et al. (2015) found that people unresponsive under ketamine later reported vivid experiences. Insofar as those reports are reliable, a marker that disappeared whenever subjects stopped responding would miss conscious states.

### *8.2 Falsification conditions*

A test of the bridge requires a model independently judged complete, and complete in the sense of Section 3.3: one that carries the quantities the system itself is determining. The bridge is thus tested together with the location hypothesis and its premise, never alone, and a physicist's model that follows every signal and is regular throughout refutes nothing. A regular model of the fuller kind, of a state known to involve experience, would challenge the necessity of a singularity; a relevant singularity in a state independently known to lack experience would challenge its sufficiency. Each test presupposes a way of saying what a system requires, which Section 4.3 leaves open.

A proposed physical realization can be tested sooner. A demonstrable failure of a system's own quantification in a state in which that system is independently known to lack experience would count against the bridge and the location hypothesis taken together. So would experience in a state in which all the system's quantitative requirements could be met. For a person the system is the one defined in Section 6.1, and a failure elsewhere in the brain bears on neither test. Ruling out one proposed neural failure, however, leaves others open. QHT gives no general way to exhaust them, though repeated unsuccessful searches would weaken the location hypothesis.

The further layers can fail separately. A person's experience marked by a failure of some other form would refute the loop hypothesis and leave the bridge and the location hypothesis standing. Experience in a brain that obeyed detailed balance at every scale that could carry a loop would rule out the circuits proposed in Section 6.2, though not the loop hypothesis itself (Section 4.4), and until a candidate circuit is specified that test is blunt. Likewise, a model might correctly mark the presence of experience and yet fail to predict which colors a person sees or which features appear together. That would count against the proposed law for character or binding, not against the bridge.

### *8.3 What an experiment would have to show*

An empirical theory of consciousness should offer formal precision, empirical grounding and explanatory productivity (Kleiner, 2020). Three requirements bear on any test of QHT.

The first is to examine internal causal structure, as the unfolding objection of Doerig et al. (2019) brings out. Systems that give identical reports may differ in whether their own quantification fails in principle, and a distinction that an investigator cannot recover from a noisy neural recording may still be available to the circuit itself.

The second is to distinguish the existence of singular states from the occupation of one. A system may approach a horizon without crossing it, or never reach it at all. To compare sleep, anesthesia and waking, a model would have to establish which happened in the episode for which we have independent grounds to judge the subject's experience.

The third is intervention, where QHT and the rival of Section 3.2 make opposite predictions. Suppose we could repair every failure of a person's own quantification, in the sense of Section 6.1, cutting each frustrated loop at a single link, while preserving the ability to distinguish blue from green, and with it the successful visual processing in which the rival account locates the experience of a blue patch. The bridge and the location hypothesis together imply that the person would cease to have any experience. On the rival account, the person would go on seeing blue.

Narrower repairs test narrower claims, and reports must be read with care. Suppose a failure in one auditory process has been linked, on independent grounds, to hearing a tone, and that process is repaired. If the listener stops reporting the tone, the repair may only have removed a cue used in reporting it, and we would need independent grounds for thinking otherwise. If the listener goes on reporting it, the report counts against the proposed link, unless we can explain why an ordinarily trustworthy judgment fails under this intervention.

## 9 Relationship to Existing Theories

### *9.1 The substrate question*

Could a machine made of different materials have the same experiences as a person? Computational functionalism holds that any system implementing the right computations is conscious, whatever its physical medium. Chalmers (1996) holds that preserving a subject's fine-grained functional organization preserves its experience. Seth's (2026) biological naturalism gives a constitutive role to the processes by which living systems maintain themselves.

QHT leaves the answer open. The issue is which features of a physical system are responsible for the failure of quantification. A horizon grounded entirely in how signals interact could carry over to different hardware. One that depended on a property specific to living cells would require that property as well. The loop hypothesis is more definite than the bridge. The test of Section 4.3 asks only what a system's signals do, so a loop that cannot close would mark experience in any material; but parts of the machine must themselves be on it, and a machine that only works out what the loop would do is not (Section 6.3). On this hypothesis living matter is not required, and producing the right results is not enough.

### *9.2 Points of contact and divergence*

QHT can complement a theory of how perceptual contents become available for use. Where two accounts give different answers about whether anything is felt at all, they also compete.

*Integrated Information Theory (IIT).* Tononi's theory starts from features of experience: a scene has distinguishable parts, yet we see it as a whole. IIT identifies the experience with the intrinsic structure of cause and effect within the system (Albantakis et al., 2023). Lee-Youngzie et al. (2026) likewise ask how local qualities fit together in one experience, using binocular rivalry to illustrate how the two eyes' views can fail to form a single stable image. QHT shares that aim. Its further question is where a model of that structure could register a quality whose intrinsic character cannot be quantified, and the loop hypothesis answers that the failure to fit is itself the register.

*Global Workspace Theory (GWT).* In the global neuronal workspace account, Dehaene and colleagues associate conscious access with the amplification and widespread sharing of a signal (Mashour et al., 2020). Lamme's (2006) recurrent processing theory gives local feedback a role in what we see even before we can report it. Either process could contribute to a horizon, and QHT agrees with Lamme that loops matter. The accounts diverge from QHT wherever broadcast or feedback is held sufficient for experience even when the brain can meet all its quantitative requirements: a loop that closes is as regular as a single sweep.

*Predictive Processing.* Predictive accounts explain how the brain revises a model in light of errors (Lee and Mumford, 2003; Friston, 2010). An error is not a horizon. If we reach for a cup where we usually leave it and see it on another shelf, we correct the expectation, and the cup's new location supplies a perfectly good answer. A horizon would arise only if the visual system required a position consistent with the information it was using and no position met that requirement: an error that no correction could cancel.

*Higher-Order Theories.* Rosenthal's (2005) account holds that mental states are conscious when suitably represented at a higher order. But why should representing a state make it feel like something? QHT's answer is that representation is not what matters. A system that represents its own states and acts on the result can thereby be on a loop (Section 4.3), and where that loop cannot close, quantification cannot succeed.

*Strange loops.* The nearest relative of the loop hypothesis is Hofstadter's (2007) proposal that the self is a "strange loop." A mind that represents itself, on this account, climbs through levels of description and arrives back where it began, and the 'I' is what the circuit produces. The shape is that of the loops of Section 4.2, and QHT shares the conviction that a subject is to be sought in a loop and in none of its parts. It differs on three points. Hofstadter's loop is a pattern among symbols, a matter of what a mind represents; QHT's lies in what a physical system's signals do, whatever they represent (Section 6.3). Hofstadter's loop succeeds: self-reference is achieved, as in Gödel's sentence, and the self is its product. QHT's fails: what marks experience is that the circuit cannot close. And Hofstadter regards the resulting 'I' as a mirage and has no place for qualia, where QHT takes both to be real. Humphrey (2011, 2023) also models consciousness on the Penrose triangle and traces sensation to a feedback loop; but there the triangle stands for an illusion, and the loop settles into an attractor. The loop hypothesis asks of a loop what neither account does: that it cannot close.

*Dynamical accounts.* Work that ties consciousness to brain activity that never settles (Rabinovich and Varona, 2018; Sanz Perl et al., 2021) describes from outside what the loop hypothesis describes from inside, where an unmet comparison must move some signal (Section 4.4). QHT adds a reason for the connection, and a condition that not all restless activity meets: it must carry comparisons of the system's own that cannot all be satisfied together.

*Non-algorithmic Approaches.* Penrose (1989) appealed to Gödel's theorem in arguing that human mathematical understanding exceeds the reach of algorithms. QHT looks for a different kind of limit. The calculation by which a scientist establishes that a system's own quantification has failed can be elementary.

*Judgments about experience.* Chalmers (2018) draws attention to our sense that experience itself is hard to explain. A listener who detects a conflict between two estimates of a sound has reason to say they disagree, but the judgment that no number could convey how the sound feels would need a separate explanation. Dennett (1991) challenges the idea that once we have explained how a listener distinguishes, remembers and describes a sound, a further intrinsic quality remains to be explained. For the illusionist, introspection represents us as having intrinsic phenomenal properties we lack (Frankish, 2016). QHT takes such qualities to be real. Both views allow a physical explanation of how we reach judgments about experience, and if they predict the same report through the same mechanism, the report alone cannot decide between them.

## 10 Philosophical Implications

### *10.1 The hard problem, reframed*

The hard problem asks how physical processing gives rise to experience (Chalmers, 1995). It has been pursued under the assumption that a complete physical description would be a complete quantitative description—that the language of physics exhausts the language of the physical. QHT challenges that assumption.

The quantitative description is not the whole physical story. The hard problem takes a particular form when the methodological exclusion of qualities from physics is reinterpreted as an ontological claim that the physical world contains only quantities—leaving consciousness apparently non-physical.

Once we accept that reality has both quantifiable and unquantifiable aspects, the question changes. Where in a quantitative model could traces of the unquantifiable be found? QHT proposes that they are found at singularities. A singularity marks the place of the quality in a description; the quality itself belongs to the physical system the description is about.

Much can be investigated while the deeper question stays open (cf. Majeed, 2016): when a color is seen, how its shape and position are experienced by the same subject, how seeing it guides action. Laws connecting experience to physical organization, of the kind Chalmers (1995, 1996) envisages, might tell us all this, even which shade a person sees. The reference to that shade would still come from someone's acquaintance with it.

### *10.2 Classic thought experiments*

*What is it like to be a bat?* Nagel (1974) asks what it means to occupy another creature's point of view. A complete physical account of a bat might establish that there is something it is like to be one, and tell us which sensory distinctions the bat makes and how its experiences change as it moves. All of this would still be knowledge of the animal from outside; reading the account does not give us its point of view.

*The Chinese Room.* Searle (1980) imagines a person who understands no Chinese applying rules to Chinese symbols, so that the room as a whole produces appropriate answers in Chinese, and asks whether following the rules suffices for understanding. QHT raises a further question, about experience: the person inside is conscious, but is the room-as-system a further subject? Neither the person's lack of understanding nor the room's appropriate answers settle that. On QHT it turns on whether person and room together realize a further system whose own quantification fails in principle.

*Mary's room.* Jackson's (1982, 1986) Mary knows all the physical facts about color but has never experienced color. QHT understands this as exhaustive quantitative knowledge; whether that exhausts reality is precisely what the thought experiment puts at issue. Mary might even know where a horizon lies in the visual system and which state seeing red will bring about. On seeing red, Mary gains an experience that can be identified as "this quality," and with it an anchor for relating other qualities to their physical conditions. The model could anticipate this acquisition without conveying what is acquired.

*Zombies.* A philosophical zombie is a physical duplicate of a conscious person who nevertheless has no experience (Chalmers, 1996). In a world governed by QHT's bridge, duplicating all the physical facts about a conscious person would also duplicate the organization that marks experience, so the duplicate would feel something. The bridge alone leaves open what it would feel, whether the correspondence is an identity or a contingent law, and whether another world could lack that law. Duplicating only a person's behavior or computation leaves more open still (Section 6.3).

*Fading and dancing qualia.* Chalmers (1996) imagines replacing neurons with components that preserve their fine-grained functional roles. If the subject's experience faded or changed during replacement, even the judgments by which the subject recognizes such changes would stay the same. QHT predicts that experience remains present wherever the replacements preserve the relevant horizon. On the loop hypothesis they would wherever the replacements are themselves parts that compare signals as the neurons did, whatever they are made of and whether or not a clock times them (Section 6.3). If the replacements eliminated every relevant failure while preserving those judgments, the account would have to identify the physical change responsible and explain how the judgments survived the loss of experience. Predicting the disappearance of a particular quality while other experience continues would require a further link to the process that was replaced.

*The inverted spectrum.* Suppose two people make the same color discriminations and associations. Could the colors nevertheless look different to them? Structural fidelity (Section 7.4) narrows the possibilities. If red and orange are neighbors in the first palette, the qualities that take their places in the second must also be neighbors, and an inversion would have to preserve every such relation. Anchors—independently established pairings between brain states and known qualities—and laws of change or combination could narrow the possibilities further. Any rearrangement that survived them all would remain open.

### *10.3 Metaphysics*

QHT's metaphysics is a dual-aspect monism—non-being contrasted against two aspects of being, the quantifiable and the phenomenal. The ontological triptych gives this precise structure, anchoring each mode of being to its signature in formal description. It could also be called a restricted property dualism. Either way, the pain in an injured limb and the neural activity associated with the pain belong to one person, and the brain signals that represent the limb are themselves part of that physical system. There is no need for a separate subject to receive the pain or a further force to make it felt.

Chalmers (1996, p. 129) expressed a related possibility: "Perhaps the physical and the phenomenal will turn out to be two different aspects of a single encompassing kind, in something like the way that matter and energy turn out to be two aspects of a single kind." QHT gives this thought a selective form. The same system may have only quantitative aspects in one state and both quantitative and phenomenal aspects in another.

Familiar cases of emergence show how a whole can have quantitative properties that its parts lack in isolation. The claim here goes further: an organized whole can have a quality whose intrinsic character no magnitude conveys. The bridge proposes when that happens. What makes this whole, with these boundaries, one subject is a further question, and the loop hypothesis suggests an answer: a loop has a definite extent, and its failure belongs to it entire (Sections 5.4 and 6.1).

This selective account takes up Russell's (1927) suggestion that a description of structure may leave intrinsic nature unspecified, and locates the qualitative aspect in particular organizations of matter. Kant (1781/1787) distinguished appearances from things considered in themselves. For QHT, the pain a person feels is directly known in experience and belongs to a system in the empirical world. Science could investigate where it occurs.

Husserl used the word *horizon* for aspects of the world implied in an experience, such as the far side of a cup or the continuation of a melody (1950/1960, §19), and traced how Galileo's mathematical conception of nature set aside its lived qualities (1936/1970). The term takes a different direction here. Husserl's horizon belongs to experience and points toward the world. QHT's belongs to a description of the world and points toward experience.

Wittgenstein brings out a related point through the eye and the visual field in the *Tractatus* (1921/1922, 5.632–5.633, 5.641). We see from a point of view while the eye itself is absent from what we see. Wittgenstein places the metaphysical subject at the limit of the world and says that the inexpressible shows itself (6.522). A QHT horizon, by contrast, can itself be stated: a scientist could describe exactly where quantification fails, even though the quality would remain unexpressed.

The *Tractatus* ends, "Whereof one cannot speak, thereof one must be silent" (7). QHT suggests that mathematics could show where that silence falls. It gives mathematical form to an old apophatic idea: that tracing the limits of what can be said can tell us something about what lies beyond them.

The later Wittgenstein (1953, §304) wrote of sensation:

> *it is not a something, but not a nothing either!*

Wittgenstein was making a grammatical point about sensation, but the words also express the intuition behind the triptych. A quale is not a something—a quantifiable aspect of the world that corresponds to a value—but neither is it a nothing, an absence that corresponds to zero. QHT proposes that a quale is an unquantifiable aspect of the world whose presence corresponds to a singularity.

A more unexpected connection comes from the life of Srinivasa Ramanujan, who attributed mathematical insights in dreams to the goddess Namagiri. Ramanujan's biographer describes the dreams that came during an illness in England (Kanigel, 1991, pp. 36, 291).

> *At one point, he had nightmares in which he was visited by images of his own abdomen as a kind of mathematical appendage with "singularities," points in space marked by indefinable mathematical surges… Intense pain might show up at $x = 1$, half as much pain at $x = -1$, and so on.*

An anecdote is not an argument. But this one is strikingly well-formed in QHT's terms. It is a representation of the dreamer's own body—a self-model. The relations of intensity among the loci are quantifiable—twice the pain here as there—even as the pain itself, at each locus, is not a value the model assigns. When so fluent a mathematical imagination represented its own suffering to itself, singularities gave the pain an address.

## 11 What QHT Achieves and What Remains Open

### *11.1 Achievements*

The paper's main theoretical proposal is that singularities—places where quantification fails in principle—could mark the physical states in which qualia arise. The further proposal, that they lie in a system's own quantitative activity, makes the idea concrete enough to investigate, and the loop hypothesis says what to look for there: comparisons that cannot close. It also says what an observer would see wherever an unmet comparison must move some signal: activity that cannot settle, of a kind that pervades the working brain. The accounts of ineffability, privacy, subjectivity, and unity follow a single line of thought: the mark is open to an observer; what it marks belongs to the system itself. And if the same physical organization could be built into a machine, the machine might have experiences of its own.

### *11.2 Open questions*

The open questions divide along the lines drawn in Section 2.4: Category A, the structural questions the theory addresses and must therefore eventually answer in full, and Category B, the questions about intrinsic nature.

The first task under Category A is the one flagged in Section 4.3. One physical system can be described in many ways, and showing that a failure survives a change of coordinates still leaves us to establish which quantities the system itself uses. In a brain we must identify what its own signals are required to supply, independently of any verdict about experience, and only then ask whether their limits occur where experience is independently known to be present, ordinary waking perception included (Section 6.2). A worked example in a model cannot by itself decide that question. Nor can it say which circuits make up the person whose experience is in question (Section 6.1). And beneath these lies the question of what makes a condition that a system imposes on its own results a requirement at all (Section 4.3).

Two further tasks are concrete. Section 7 offers one way in which several physical failures might mark a single structured experience: could we predict, from the grouping of the failures alone, which features a person sees as belonging to the same object? And the system whose activity moves the hand is the one that feels the pain: could the withdrawal be explained specifically by how the pain feels?

Within Category B, the palette question admits an indirect approach through the relations among qualities (Section 7.4). The qualitative character question cuts deeper. Why does experience have these specific characters—color, pain, sound—rather than some entirely other set we cannot conceive? A deeper theory might explain why a particular quality accompanies a particular physical state even if its equations could never convey what the quality is like. The existence question—why reality has unquantifiable aspects at all—remains open independently, on a par with why there is something rather than nothing.

QHT has a limited affinity with mysterianism, the view that our powers of understanding set limits to an explanation of consciousness (McGinn, 1999). The limit proposed here lies elsewhere: it concerns what quantitative description can convey. Even if we found a law connecting particular brain states to particular qualities, we could still ask why nature obeys it. To adapt Hawking (1988), what "breathes fire into the equations" and gives them qualia to account for remains unknown.

### *11.3 Closing*

In *Dreams of a Final Theory*, Weinberg (1992) described what it would mean to make progress on this difficulty.

> *Suppose … that we will come to understand the objective correlatives to consciousness in terms of physics… It is not unreasonable to hope that when the objective correlatives to consciousness have been explained, somewhere in our explanations we shall be able to recognize something, some physical system for processing information, that corresponds to our experience of consciousness itself, to what Gilbert Ryle has called "the ghost in the machine." That may not be an explanation of consciousness, but it will be pretty close.*

Weinberg hoped that somewhere in our model of the physical correlates of consciousness we would recognize something that corresponds to consciousness itself. QHT proposes that singularities could fill this role: the equations could show where a quality is present, even though only the subject knows it from within. If QHT is correct, it would not be a complete explanation of consciousness. But by Weinberg's standard, it would be pretty close.

## Acknowledgements

The author consulted Claude Fable 5.2 (Max) by Anthropic, ChatGPT 6.2 Astra (Ultra) by OpenAI and Gemini 4 Neon (High) by Google DeepMind in the development of this paper. The author takes full responsibility for all content.